\RequirePackage[2020-02-02]{latexrelease}
\documentclass
[aps,preprint,showkeys,a4paper,tightenlines,superscriptaddress]{revtex4}%
\usepackage{eurosym}
\usepackage{amsfonts}
\usepackage{amsmath}
\usepackage{amssymb}
\usepackage{graphicx}
\usepackage{subfig}
\usepackage{caption}%
\providecommand{\U}[1]{\protect\rule{.1in}{.1in}}
\RequirePackage[2020-02-02]{latexrelease}
\RequirePackage[2020-02-02]{latexrelease}
\RequirePackage[2020-02-02]{latexrelease}
\RequirePackage[2020-02-02]{latexrelease}
\RequirePackage[2020-02-02]{latexrelease}
\RequirePackage[2020-02-02]{latexrelease}
\RequirePackage[2020-02-02]{latexrelease}
\RequirePackage[2020-02-02]{latexrelease}
\providecommand{\U}[1]{\protect\rule{.1in}{.1in}}

\begin{document}
\title{Theory of viscoelastic adhesion and friction}
\author{G. Carbone}
\affiliation{Corresponding author. Email: giuseppe.cabone@poliba.it}
\affiliation{Department of Mechanics, Mathematics and Management, Politecnico of Bari, V.le
Japigia, 182, 70126, Bari, Italy}
\affiliation{Imperial College London, Department of Mechanical Engineering, Exhibition
Road, London SW7 2AZ}
\affiliation{CNR - Institute for Photonics and Nanotechnologies U.O.S. Bari, Physics
Department \textquotedblright M. Merlin\textquotedblright, via Amendola 173,
70126 Bari, Italy}
\author{C. Mandriota}
\affiliation{Department of Mechanics, Mathematics and Management, Politecnico of Bari, V.le
Japigia, 182, 70126, Bari, Italy}
\author{N. Menga}
\affiliation{Department of Mechanics, Mathematics and Management, Politecnico of Bari, V.le
Japigia, 182, 70126, Bari, Italy}
\affiliation{Imperial College London, Department of Mechanical Engineering, Exhibition
Road, London SW7 2AZ}
\keywords{viscoelasticity, adhesion, friction, crack propagation, hysteresis}
\begin{abstract}
We present a novel theory of the adhesive contact of linear viscoelastic
materials against rigid substrates moving at constant velocity. Despite the
non-conservative behavior of the system, the closure equation of the contact
problem can be rigorously formulated in the form of a local energy balance. In
the case of adhesiveless contacts, this is equivalent to enforce the
stationarity of the total energy stored into the viscoelastic material.
However, in the presence of interfacial adhesion, the appearance of
non-conservative terms leads to different values of the energy release rates
G1 and G2 at the contact trailing and leading edges, respectively.
Specifically, the present theory predicts a non-monotonic trend of G1 and G2
as function of the indenter velocity, as well as a very significant
enhancement of hysteretic friction due to the coupling between adhesion and
viscoelasticity, compared to the adhesiveless case. Both predictions are in
very good agreement with existing experimental data.

\end{abstract}
\maketitle

\section{Introduction}

Understanding the origin of friction in adhesive sliding or rolling contact of
rubber-like materials is a long-standing problem. Recent theories
\cite{Persson2001,Persson2004,Menga2014,Menga2021} have shown that part of the
friction is ascribable to dissipative phenomena occurring in the bulk of the
material, associated with the intrinsic viscoelastic behavior of natural
rubbers and elastomers. However, experimental observations made by Grosch in
his seminal paper \cite{Grosch1963}, have demonstrated the key role that
interfacial adhesion plays in determining the overall frictional response in
sliding contacts \cite{LeGal2005}. Nonetheless, although a comprehensive
theory of adhesive viscoelastic friction is still lacking, the adhesive
contribution to friction is commonly postulated in describing the frictional
behavior of both micro- and macro-scale systems
\cite{Kim2007,Zhou2013,Sharp2016}. Adhesive layers adsorption
\cite{Yoshizawa1993_1,Yoshizawa1993_2,Bhushan1995,Maeda2002}, and the dynamics
of adhesive links debonding/rebinding
\cite{Chernyak1986,Chaudhury1996,Vorvolakos2003} are some of the mechanisms
responsible for adhesive hysteresis even in the case of rigid and purely
elastic interfaces, which eventually lead to a friction force opposing the
interface sliding. However, regardless of the detailed description of the
adhesive links, specific scale-dependent mechanisms of interaction between
adhesion and viscoelasticity can be identified. Indeed, at the large-scale,
the effect of adhesion is to increase the real contact area and, in turn, the
portion of bulk material undergoing cyclic deformation, which eventually leads
to the aforementioned bulk energy dissipation and to viscoelastic friction
\cite{Scaraggi2015}. On the other hand, at the asperity level, as observed in
laboratory tests
\cite{barquins1978,Charmet1996,Zhang2015,roberts1979,Hoyer2022}, local
adhesion hysteresis modifies the energy release rates at the leading and
trailing contact edges thus producing an additional contribution to friction,
usually referred to as adhesive friction \cite{She1998,Hao2007,LeGal2005}.
This mechanism, usually ascribed to the so-called small-scale viscoelasticity
\cite{Baney1999,Persson2005,Persson2021}, is triggered by adhesion (i.e., it
vanishes for adhesiveless contacts), and leads to an increase of the contact
area and an asymmetric distribution of contact stresses \cite{Carbone2004}. In
this letter we present a novel theory, based on energy balance, to study
adhesive viscoelastic contacts in steady sliding motion. The proposed approach
allows to investigate the interplay between adhesion and viscoelasticity
across the scales for speed values spanning the entire viscoelastic spectra of
the material. In agreement with experimental evidences
\cite{Grosch1963,roberts1979,Zhang2015}, we predict, for the first time, a
specific velocity-dependent frictional behavior, which proves that the
adhesive contribution to viscoelastic friction cannot be neglected.

\section{Formulation}

We consider the case of a rigid periodic indenter, under displacement
controlled conditions, in adhesive contact with a linear viscoelastic
half-space. The indenter moves at a constant velocity $\mathbf{v}$. According
to \cite{Carbone2013,Menga2014}, the surface normal displacement $u\left(
\mathbf{x}\right)  $ and stress $\sigma\left(  \mathbf{x}\right)  $ fields (in
a reference frame co-moving with the indenter) are time-independent and
related to each other by a spatial convolution product
\begin{equation}
u\left(  \mathbf{x}\right)  =\int d^{2}x_{1}\mathcal{G}_{\mathbf{v}}\left(
\mathbf{x-x}_{1}\right)  \sigma\left(  \mathbf{x}_{1}\right)  \label{eq1}%
\end{equation}
where, $\mathbf{x}$ is the in-plane position vector, and $\mathcal{G}%
_{\mathbf{v}}$ is the viscoelastic Green's function, which parametrically
depends on $\mathbf{v}$. Notably, $\mathcal{G}_{\mathbf{v}}\left(
\mathbf{x}\right)  $ is non-symmetric, i.e. $\mathcal{G}_{\mathbf{v}}\left(
-\mathbf{x}\right)  \neq\mathcal{G}_{\mathbf{v}}\left(  \mathbf{x}\right)  $
and, therefore, can be decomposed in a symmetric (even) part $\mathcal{G}%
_{\mathbf{v}}^{E}\left(  \mathbf{x}\right)  =\mathcal{G}_{\mathbf{v}}%
^{E}\left(  -\mathbf{x}\right)  $ and antisymmetric (odd)\ one $\mathcal{G}%
_{\mathbf{v}}^{O}\left(  \mathbf{x}\right)  =-\mathcal{G}_{\mathbf{v}}%
^{O}\left(  -\mathbf{x}\right)  $, i.e.%
\begin{equation}
\mathcal{G}_{\mathbf{v}}\left(  \mathbf{x}\right)  =\mathcal{G}_{\mathbf{v}%
}^{E}\left(  \mathbf{x}\right)  +\mathcal{G}_{\mathbf{v}}^{O}\left(
\mathbf{x}\right)  \label{eq decomp}%
\end{equation}
We recall that the contact area is an unknown of the problem and must be
determined as a part of the solution. To this aim we need an additional
closure equation, which can be phrased in terms of energy balance between the
change of surface energy $\Delta\gamma\delta A$, caused by an infinitesimally
small change $\delta A$ of the contact area $A$, and the corresponding work of
the internal stress $\delta L$,%
\begin{equation}
\delta L=\Delta\gamma\delta A\label{closure equation}%
\end{equation}
where $\Delta\gamma$ is the surface energy per unit area also referred to as
the Dupr\`{e} work of adhesion. Note that in the case of elastic materials
$\delta L$ equates the change of the elastic energy $\delta U$ stored into the
material. However for non-conservative systems, as in our case, $\delta
L=\delta U\ +\delta L_{P}$ where $\delta L_{P}$ is a non conservative
(positive or negative) term. Aiming at expressing $\delta U$ and $\delta
L_{P}$ in terms of $\sigma\left(  \mathbf{x}\right)  $ and $u\left(
\mathbf{x}\right)  $, we consider a quasi-static (or infinitely slow) change
$\delta u\left(  \mathbf{x}\right)  $ of the displacement field $u\left(
\mathbf{x}\right)  $. The work $\delta L$ made by internal stresses is $\delta
L=\int d^{2}x\sigma\left(  \mathbf{x}\right)  \delta u\left(  \mathbf{x}%
\right)  $. Using (\ref{eq1}) and (\ref{eq decomp}) yields%
\begin{align}
\delta L_{P} &  =\int d^{2}xd^{2}x_{1}\mathcal{G}_{\mathbf{v}}^{O}\left(
\mathbf{x-x}_{1}\right)  \sigma\left(  \mathbf{x}\right)  \delta\sigma\left(
\mathbf{x}_{1}\right)  \label{work}\\
\delta U &  =\int d^{2}xd^{2}x_{1}\mathcal{G}_{\mathbf{v}}^{E}\left(
\mathbf{x-x}_{1}\right)  \sigma\left(  \mathbf{x}\right)  \delta\sigma\left(
\mathbf{x}_{1}\right)  \label{Uel}%
\end{align}
Moreover, since $\mathcal{G}_{\mathbf{v}}^{E}\left(  \mathbf{x}\right)  $ is
symmetric, we have that $U=\frac{1}{2}\int d^{2}x\mathcal{G}_{\mathbf{v}}%
^{E}\left(  \mathbf{x-x}_{1}\right)  \sigma\left(  \mathbf{x}\right)
\sigma\left(  \mathbf{x}_{1}\right)  $; similarly, with $\mathcal{G}%
_{\mathbf{v}}^{O}\left(  \mathbf{x}\right)  $ being antisymmetric, we have
$\int d^{2}x\mathcal{G}_{\mathbf{v}}^{O}\left(  \mathbf{x-x}_{1}\right)
\sigma\left(  \mathbf{x}\right)  \sigma\left(  \mathbf{x}_{1}\right)  =0$.
Hence, from Eq. (\ref{Uel}), the elastic energy $U$ can be rephrased in a more
convenient form as%
\begin{equation}
U=\frac{1}{2}\int d^{2}x\sigma\left(  \mathbf{x}\right)  u\left(
\mathbf{x}\right)  \label{usual elastic energy}%
\end{equation}
which is the standard form already known for purely elastic materials.
Notably, since $\mathcal{G}_{\mathbf{v}}^{O}\left(  \mathbf{x}\right)  $ is an
odd function, the term $\delta L_{P}\ $vanishes when the shape of the
displacement field during the loading process does not change, i.e. for
$\delta u\left(  \mathbf{x}\right)  =$ $u_{0}\left(  \mathbf{x}\right)
~\delta\eta$, where $\eta$ is a dimensionless control parameter, and
$u_{0}\left(  \mathbf{x}\right)  $ represents the fixed shape of the
displacement field. Linearity (\ref{eq1}) yield $\sigma\left(  \mathbf{x}%
\right)  =\sigma_{0}\left(  \mathbf{x}\right)  ~\eta$ and $\delta\sigma\left(
\mathbf{x}\right)  =\sigma_{0}\left(  \mathbf{x}\right)  ~\delta\eta$ with the
condition $u_{0}\left(  \mathbf{x}\right)  =\int d^{2}x_{1}\mathcal{G}%
_{\mathbf{v}}\left(  \mathbf{x-x}_{1}\right)  \sigma_{0}\left(  \mathbf{x}%
_{1}\right)  $. A particular case leading to $\delta L_{P}=0$ occurs for
concentrated loads, e.g. when $\sigma\left(  \mathbf{x}\right)  =\delta
_{D}\left(  \mathbf{x-x}_{0}\right)  ~\eta$, i.e. when the stress distribution
is represented by a concentrated force at point $\mathbf{x}_{0}$, the quantity
$\delta_{D}\left(  \mathbf{x}\right)  $ being, in fact, the Dirac delta function.

Aiming at calculating the term $\delta L$ of Eq. (\ref{closure equation}), we
consider an infinitesimally small (and extremely slow) change $\delta A$ of
the contact area $A$, about an equilibrium condition, due to a small
perturbation of its boundary $\partial A$. In the presence of extremely
short-range adhesive forces, the contact area boundary $\partial A$ can be
regarded as the tip of a crack, and the contact area change $\delta A$ as a
small propagation (advancing or receding) of the crack itself. In such a case,
the displacement field $u\left(  \mathbf{x}\right)  $ discontinuously jumps of
a quantity $\Delta u\left(  \mathbf{x}\right)  $ as the contact domain is
subjected to the small change $\delta A$. It follows that $\Delta u\left(
\mathbf{x}\right)  =0$ for $\mathbf{x}\in A$ and $\Delta u\left(
\mathbf{x}\right)  =u_{+}\left(  \mathbf{x}\right)  -u_{-}\left(
\mathbf{x}\right)  \neq0$ for $\mathbf{x}\in\delta A$, where $u_{+}\left(
\mathbf{x}\right)  =$ $u\left(  \mathbf{x},A+\delta A\right)  $ and
$u_{-}\left(  \mathbf{x}\right)  =u\left(  \mathbf{x},A\right)  $. Recalling
that for $\mathbf{x}\notin A$ but close to the boundary $\partial A$ the
displacement field takes the form $u\left(  \mathbf{x}\right)  \propto\sqrt
{d}$, where $d$ is the distance from the boundary, one concludes that the
`quasi-static movement' of the crack occurs through a succession of single
point loadings as in the case of zipper opening or closing
\cite{Barquins1981,Shull2002}. Then, the infinitely small contact propagation
over the area $\delta A$ can be regarded as an opening (closing) displacement
field governed by a single parameter $\eta$ slowly increasing from zero to
one, i.e. $u\left(  \mathbf{x},\eta\right)  =\eta$~$\Delta u\left(
\mathbf{x}\right)  +u_{-}\left(  \mathbf{x}\right)  $, which yields $\delta
u\left(  \mathbf{x},\eta\right)  =\left(  \partial u/\partial\eta\right)
\delta\eta=\Delta u\left(  \mathbf{x}\right)  \delta\eta$. Linearity
(\ref{eq1}) between stresses and displacement then yields $\sigma\left(
\mathbf{x}\text{,}\eta\right)  =\eta~\sigma_{+}\left(  \mathbf{x}\right)  $
with $\mathbf{x}\in\delta A$ and $\sigma_{+}\left(  \mathbf{x}\right)  =$
$\sigma\left(  \mathbf{x},A+\delta A\right)  $. Note that for $\mathbf{x}%
\in\delta A$, the stress $\sigma_{-}\left(  \mathbf{x}\right)  =\sigma\left(
\mathbf{x},A\right)  =$ $0$. Then, the work $\delta L=\delta U+\delta L_{P}$
done by the internal stresses during the quasi-static crack opening (closing)
process can be simply calculated as%
\begin{align}
\delta L &  =\int_{\delta A}d^{2}x\int_{0}^{1}\sigma\left(  \mathbf{x}%
,\eta\right)  \Delta u\left(  \mathbf{x}\right)  \delta\eta\nonumber\\
&  =\frac{1}{2}\int_{\delta\Omega}d^{2}x\sigma_{+}\left(  \mathbf{x}\right)
\Delta u\left(  \mathbf{x}\right)  \label{internal work}%
\end{align}
Noting that $\int_{\delta A}d^{2}x\sigma_{+}\left(  \mathbf{x}\right)  \Delta
u\left(  \mathbf{x}\right)  =\int d^{2}x\left[  \sigma_{+}\left(
\mathbf{x}\right)  +\sigma_{-}\left(  \mathbf{x}\right)  \right]  \Delta
u\left(  \mathbf{x}\right)  $, and recalling that $\Delta u\left(
\mathbf{x}\right)  =u_{+}\left(  \mathbf{x}\right)  -u_{-}\left(
\mathbf{x}\right)  $, Eqs. (\ref{closure equation}, \ref{usual elastic energy}%
) then yield%
\begin{equation}
\delta U=\Delta\gamma\delta A-\delta L_{P}\label{closure equation 2}%
\end{equation}
where%

\begin{equation}
\delta L_{P}=-\frac{1}{2}\int d^{2}x\left[  \sigma_{+}\left(  \mathbf{x}%
\right)  u_{-}\left(  \mathbf{x}\right)  -\sigma_{-}\left(  \mathbf{x}\right)
u_{+}\left(  \mathbf{x}\right)  \right]
\end{equation}
is the non-conservative contribution to the work of internal stresses related
to the hysteretic behavior of the material. In fact, using Eq. (\ref{eq1}) and
recalling that $\mathcal{G}_{\mathbf{v}}^{O}\left(  \mathbf{x}\right)
=\frac{1}{2}\left[  \mathcal{G}_{\mathbf{v}}\left(  \mathbf{x}\right)
-\mathcal{G}_{\mathbf{v}}\left(  -\mathbf{x}\right)  \right]  $, we write
$\delta L_{P}=-\int d^{2}xd^{2}x_{1}\mathcal{G}_{\mathbf{v}}^{O}\left(
\mathbf{x-x}_{1}\right)  \sigma_{+}\left(  \mathbf{x}\right)  \sigma
_{-}\left(  \mathbf{x}_{1}\right)  $, thus showing that this contribution is
associated with the antisymmetric component of the Green function. It follows
that in the case of purely elastic materials, i.e. for symmetric Green's
functions, $\delta L_{P}=0$.

Recalling the definition of the energy release rate $G=\delta U/\delta A$ we
rewrite (\ref{closure equation 2}) in the form%

\begin{equation}
G\left(  \mathbf{v}\right)  =G_{0}-\frac{\delta L_{P}}{\delta A}%
\label{energy_equil}%
\end{equation}
where we set $G_{0}=\Delta\gamma$. Note that standard theories of viscoelastic
crack propagation \cite{Persson2005, Carbone2005,Carbone2005epje} usually
assume a wide non linear process zone at the crack tip leading to values of
$G_{0}$ greater than $\Delta\gamma$. The present theory do not include such
non linear processess as cavitation and bond breaking. However, at least for
adhesive cracks, where the tensile stress is not so high as to generate
cavitation or other ruptures, the effect of crack process zone may not be
important unless the rubber compound is very weakly crosslinked.

Interestingly, Eq. (\ref{energy_equil}) shows that the effect of the
viscoelasticity can be either to increase or decrease the energy release rate,
depending on the sign of the $\delta L_{P}$. This leads, therefore, to a
different behavior at the leading and trailing edges of the contact. We also
observe that in the case of adhesiveless contact, i.e. $\Delta\gamma=0$, the
stress intensity factor at the edge of the contact vanishes so that the
displacement field does not present any discontinuous jump as the contact area
changes of the infinitesimally quantity $\delta A$, leading to $\delta
L=\delta U=\delta L_{P}=0$. Thus, for adhesiveless steady-state viscoelastic
contacts (under displacement controlled conditions), the equilibrium solution
can be found by simply enforcing $\delta U/\delta A=0$, as in the purely
elastic case.\begin{figure}[ptbh]
\centering\includegraphics[width=0.9\textwidth]{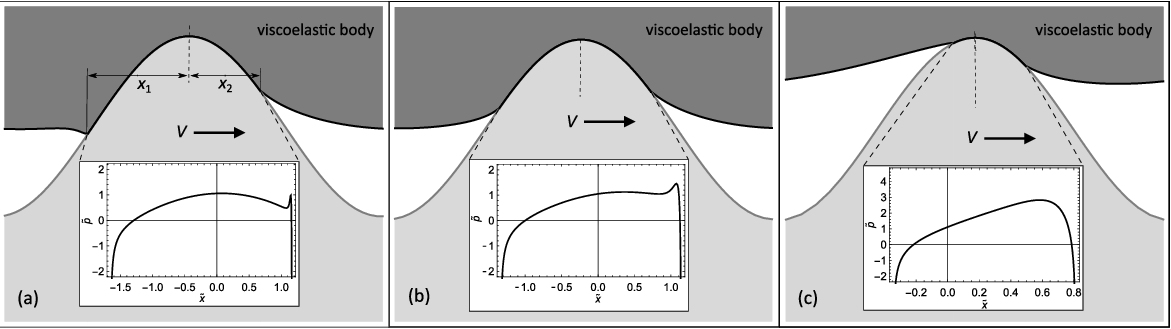}\caption{The sliding
contact of a viscoelastic material against a moving rigid indenter, in the
presence of adhesion. The inset shows the asymmetric pressure distribution at
the contact interface. The contact half-width is $a=\left(  x_{2}%
+x_{1}\right)  /2$, and contact eccentricity is $e=\left(  x_{2}-x_{1}\right)
/2$. Small-scale viscoelasticy (also referred to as adhesion hysteresis)
occurs at small velocities (regime I), (a); coupled small and large scale
viscoelasticy (regime II) occurs at intermediat velocities (b); large scale
viscoelasticity (regime III) occurs at higher velocities, (c).}%
\label{fig1}%
\end{figure}

\section{Results and Discussion}

In what follows, we comment on the viscoelastic adhesive behavior predicted
for a 1D+1D contact of a viscoelastic half-plane in partial contact with a
wavy rigid profile of wavelength $\lambda$, wavevector $k=2\pi/\lambda$ and
amplitude $h$, (see Fig. \ref{fig1}), sliding at constant velocity $v$. The
material obeys the standard linear viscoelastic model with one single
relaxation time $\tau$, and creep function $J\left(  t\right)  =1/E_{0}%
-\left(  1/E_{0}-1/E_{\infty}\right)  \exp\left(  -t/\tau\right)  $, where
$E_{0}$ and $E_{\infty}$ are respectively the low and high frequency real
values of the viscoelastic modulus of the material. The contact is modelled by
relying on Green's function approach described in Refs
\cite{Menga2016,Menga2018,Menga2021}. The closing conditions to calculate the
contact area half-width $a=\left(  x_{2}+x_{1}\right)  /2$, and the
eccentricity $e=\left(  x_{2}-x_{1}\right)  /2$, are given by Eq.
(\ref{energy_equil}). Specifically, we consider two independent contact area
variations at the two edges of the contact. Due to material viscoelasticity
the contact interface will experience an asymmetric distribution of normal
pressure, depending on the specific contact conditions. This leads to
viscoelastic friction, so that, according to Ref.
\cite{Persson2001,Persson2004,Menga2016,Menga2018,Menga2021}, the global
friction coefficient $\mu$ can be calculated as%
\begin{equation}
\mu=-\frac{1}{\lambda p_{\infty}}\int p\left(  x\right)  u^{\prime}\left(
x\right)  dx \label{frict}%
\end{equation}
where $p_{\infty}$ is the remote pressure acting on the solid and $u^{\prime
}\left(  x\right)  $ the spatial derivative of the normal displacements.\emph{
}Notably, \emph{ }the only source of contact pressure asymmetry is ascribable
to the viscoelastic energy dissipation occurring during the cyclic deformation
of viscoelastic half-space. The viscoelastic material is indeed excited at
different frequencies. The bulk of the material is excited at frequency
$\approx v/\lambda$; therefore, depending on the indenter sliding velocity
$v$, the bulk of the material may be in the rubbery, transition or glassy
regions. Close to the edges of the contact the material is excited at
frequency $v/\rho$, where $\rho$ is the velocity dependent radius of curvature
of the blunted edges of the contact (in general, $\rho$ may differ at the
trailing and leading edges). This allows us to roughly identify three regimes,
depending on the indenter sliding velocity $v$: (I) the small-scale
viscoelastic regime [Fig. \ref{fig1}(a)], for $v\leq\rho/\tau<\lambda/\tau$,
in which viscoelastic hysteresis is mostly localized close to the edges of the
contact (the so-called adhesion hysteresis) and the bulk of the material
behaves elastically with modulus $E_{0}$; (II) the coupled small/large scale
viscoelastic regime [Fig. \ref{fig1}(b)], for $\rho/\tau<v<\lambda/\tau$, in
which the whole solid experiences viscoelastic hysteresis; (III) the bulk
viscoelastic regime [Fig. \ref{fig1}(c)], for $\rho/\tau<\lambda/\tau\leq v$,
in which viscoelastic hysteresis mostly affects the bulk material, whereas the
response of the material close to the edges of the contact is elastic, with
modulus $E_{\infty}$. Notice that for $v\ll\rho/\tau$ and $\lambda/\tau\ll v$
the whole half-plane behaves elastically, with Young moduli $E_{0}$ and
$E_{\infty}$, respectively. Regarding the frictional response, in the
small-scale viscoelastic regime (i.e. regime I), the friction coefficient is
dominated by adhesion hysteresis and we get $\mu\approx\mu_{a}=\left(
G_{1}-G_{2}\right)  /\lambda p_{\infty}$. According to \cite{Carbone2004} we
refer to $\mu_{a}$ as the adhesive friction coefficient. In the bulk
viscoelastic regime (i.e. regime III), we expect a vanishing adhesive
contribution to friction (i.e., $G_{1}\approx G_{2}\approx\Delta\gamma$), as
most of the energy dissipation occurs into the bulk of the solid, so that
$\mu\approx\mu_{0}$, with $\mu_{0}$ being the adhesiveless friction
coefficient. Interestingly, regime II\ is characterized by a strong coupling
of small and large scale viscoelastic hysteresis, which leads to a very strong
enhancement of friction due to the combined effect of adhesion hysteresis and
bulk viscoelasticity.

\begin{figure}[ptbh]
\centering\includegraphics[width=0.5\textwidth]{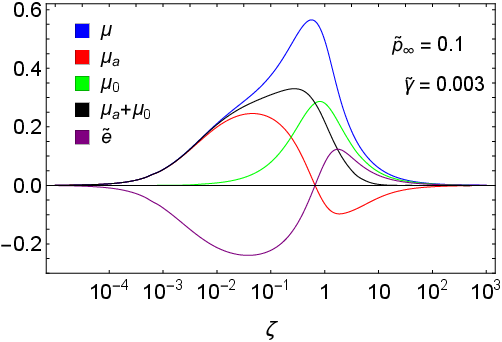}\caption{The
frictional response under load controlled conditions. $\mu$ is the overal
friction coefficient for the viscoelastic adhesive contact here considered,
$\mu_{a}$ is the adhesive friction coefficient, and $\mu_{0}$ the
corresponding adhesiveless viscoelastic friction coefficient. $\mu$ is always
larger than $\mu_{a}+\mu_{0}$, thus showing that friction calculations based
on two-scale approaches, i.e. made by summing up the adhesive friction
($\mu_{a}$) and the adhesiveless viscoelastic friction ($\mu_{0}$), does not
predict correctly the overall frictional behavior of adhesive vicoelastic
interfaces.The quantity $\tilde{e}$ is the dimensionless eccentricity of the
contact area.}%
\label{fig2}%
\end{figure}

Fig. \ref{fig2} reports the friction coefficients $\mu$, $\mu_{a}$, $\mu_{0}$,
$\mu_{a}+\mu_{0}$, and the dimensionless eccentricity $\tilde{e}=ke$ as
functions of the dimensionless sliding speed $\zeta=kv\tau$, at given
dimensionless pressure $\tilde{p}_{\infty}=2p_{\infty}/(E_{0}^{\ast}kh)=0.1$,
with $E_{0}^{\ast}=E_{0}/\left(  1-\nu^{2}\right)  $, Poisson ratio $\nu=0.5$,
$E_{\infty}/E_{0}=10$, and dimensionless adhesion energy $\tilde{\gamma
}=2\gamma/\left(  E^{\ast}\lambda\right)  =0.003$. Fig. \ref{fig2} clearly
confirms that at small values of the dimensionless speed $\zeta$, the dominant
contribution to friction arises from adhesion hysteresis, i.e., $\mu\approx
\mu_{a}$ (small-scale viscoelasticity, regime I). In the intermediate range of
speeds (regime II), the real frictional behavior is strongly enhanced by the
combined presence of adhesion hysteresis and bulk viscoelasticity, which leads
to an overall friction coefficient $\mu$ significantly greater than the sum of
the adhesive friction coefficient $\mu_{a}$ and the adhesiveless friction
coefficient $\mu_{0}$. In such conditions, the value of the friction
coefficient $\mu$ may rise up to almost two times the values $\mu_{a}+\mu_{0}%
$. The latter result confirms the fundamental role of adhesion in determining
the frictional performance of viscoelastic materials, as for example in the
case of grip and handling performance of racing tires \cite{Sharp2016}. We
stress that, in regime II, there is no possibility to separate the small and
the large scales viscoelastic losses. Models based on such an assumption fall
short in predicting the frictional and adhesive behavior of polymeric
materials in the intermediate range of sliding velocities, as observed by
Grosch in Ref. \cite{Grosch1963} (see Figs. 7-10 therein), and also pointed
out by Roberts \cite{roberts1979}. In agreement with Grosch's observations,
the trend of the overall friction coefficient $\mu$ (Fig. \ref{fig2}),
presents a hump localized where the maximum adhesive friction $\mu_{a}$ occurs
(low velocity side). As the velocity is increased, a maximum of $\mu$ is then
observed at dimensionless speed $\zeta$ close to $1$, at which the
adhesiveless friction coefficient $\mu_{0}$ also takes its maximum. We note
that the contribution of adhesion to friction can also occur from processes
related to molecules binding-stretching-detaching cycles \cite{tiwari}, which
are not included in the present treatment.

\begin{figure}[ptbh]
\centering\includegraphics[width=0.5\textwidth]{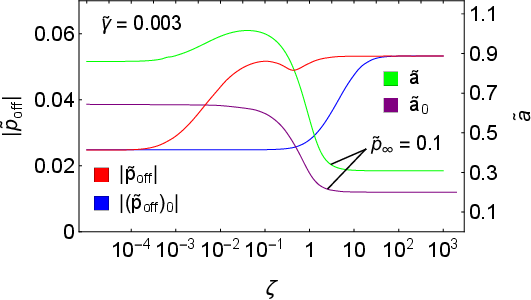}\caption{The pull-off
load $\left\vert \tilde{p}_{off}\right\vert $, the pull-off load $\left\vert
\left(  \tilde{p}_{off}\right)  _{0}\right\vert $ (see text for a detailed
description), the contact area $\tilde{a}$ and the adhesiveless $\tilde{a}%
_{0}$ contact area (at given contact pressure) as functions of the
dimensionless speed $\zeta$. At intermediate speeds, the pull-off load and the
contact area are strongly enhanced due to a strong increase of the effective
adhesion caused by viscoelastic dissipation.}%
\label{fig3}%
\end{figure}Fig. \ref{fig3} reports the dimensionless half width $\tilde
{a}=ka$ of the contact area, the dimensionless width $\tilde{a}_{0}=ka_{0}$ of
the corresponding adhesiveless viscoelastic contact, the critical remote
pull-off load $\left\vert \tilde{p}_{off}\right\vert $ and the pull-off load
$\left\vert \left(  \tilde{p}_{off}\right)  _{0}\right\vert $. Note that the
latter has been estimated assuming a perfect elastic material with elastic
modulus $\left\vert E\left(  \omega=\zeta/\tau\right)  \right\vert $. All
these quantities are plotted versus the dimensionless speed $\zeta$. By
comparing the trend of the quantity $\tilde{a}$ with $\tilde{a}_{0}$, we note
that, in contrast with the monotonically decreasing trend of $\tilde{a}_{0}$,
the half-width $\tilde{a}$ presents a significant non-monotonic behavior
characterized by a very strong enhancement of the contact area. Specifically,
such a feature occurs in the same range of intermediate speeds (regime II)
corresponding to the aforementioned increase of $\mu$ well above the the
values of $\mu_{a}+\mu_{0}$. This considerable increase of the contact area
takes place as a consequence of the combined effect of adhesion hysteresis and
bulk viscoelasticity. Coming to the pull-off remote pressure, we observe that
$\left\vert \left(  \tilde{p}_{off}\right)  _{0}\right\vert $ monotonically
increase because of the monotonic material stiffening (\ref{fig3}) occurring
as $\zeta$ is increased. However, the effect of adhesion hysteresis and
viscoelasticity causes the real pull-off load $\left\vert \tilde{p}%
_{off}\right\vert $ to rise up much beyond $\left\vert \left(  \tilde{p}%
_{off}\right)  _{0}\right\vert $, already at very low velocities (regime I). A
further increase of $\zeta$ leads to a slightly non-monotonic trend of
$\left\vert \tilde{p}_{off}\right\vert $, likely owed to the combined effect
of adhesion hysteresis and bulk viscoelasticity (regime II). Experimental
observations on rolling cylinders in contact with viscoelastic soft substrates
strongly support this picture \cite{Charmet1996}.

\begin{figure}[ptbh]
\centering\includegraphics[width=0.5\textwidth]{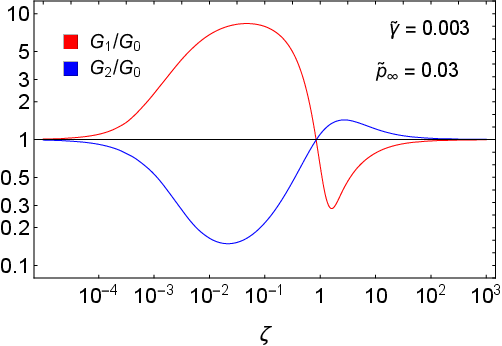}\caption{The energy
release rates $G_{1}/G_{0}$ and $G_{2}/G_{0}$ at the trailing and leading
edges respectively, as a functin of the dimensionless sliding velocity $\zeta
$, in a log-log diagram. }%
\label{fig4}%
\end{figure}To close the discussion we present in Fig. \ref{fig4}.\ the trend
of the energy release rates $G_{1}$, at the contact trailing edge, and $G_{2}%
$, at the contact leading edge, as a function of the dimensionless sliding
speed $\zeta$ in a log-log diagram. Note that, $G_{1}$ and $G_{2}$ follow a
non-monotonic trend. More specifically $G_{1}/G_{0}$ (red curve), at very low
speed takes the unit value as the material behaves elastically (with modulus
$E_{0}$) everywhere and the energy release rate must equate the adhesion
energy for unit area $G_{0}=\Delta\gamma$. As the velocity is increased,
$G_{1}/G_{0}$ increases up to a maximum value $\approx$ $E_{\infty}/E_{0}$
\cite{Greenwood2004, Persson2005, Carbone2005,Carbone2005epje}, and then
decreases to values even smaller that those attained by $G_{2}/G_{0}$,
followed by a further increase which asymptotically leads to $G_{1}%
/G_{0}\rightarrow1$ at very high velocity. Indeed, in this case, the material
is again elastic everywhere (with modulus $E_{\infty}$). Regarding
$G_{2}/G_{0}$ (blue curve), it takes the unit value at very low velocity, and
with $\zeta$ increasing, it firstly decreases to a minimum value and then
increases again to a local maximum. At very high speed, $G_{2}/G_{0}$
asymptotically decrease toward the unit value. This non-monotonic behavior is
not observed in infinite systems \cite{Greenwood2004, Persson2005,
Carbone2005,Carbone2005epje} and is a characteristic of the finiteness of
contact area as inferred in \cite{Persson2017, Persson2021}, where a bell
shaped behavior of $G_{1}$ is reported.

\section{Conclusions}

We present a comprehensive theory of the adhesive contact mechanics of
linearly viscoelastic materials in steady contact with rigid substrates moving
at constant velocity. The theory allows to investigate the interplay between
viscoelastic hysteresis and interfacial adhesion over a wide range of
velocity, thus encompassing several characteristic length scales and
viscoelastic regimes. The present theory clearly indicates that adhesion plays
a fundamental role in determining the overall frictional response of the
contact over the entire range of sliding speeds. Specifically, we show that,
at intermediate speeds, the concurrent presence of small scale viscoelasticity
(i.e. adhesion hysteresis) and large-scale viscoelasticity (bulk hysteresis)
leads to an enhancement of the overall friction coefficient $\mu$ which has
been only observed experimentally and \ never explained before. Our predicted
friction versus sliding velocity presents a strong agreement with experimental
observations reported by Grosch in his seminal work \cite{Grosch1963}.

\begin{acknowledgments}
This work was partly supported by the Italian Ministry of Education,
University and Research under the Programme \textquotedblleft Progetti di
Rilevante Interesse Nazionale (PRIN)\textquotedblright, Grant Protocol
2017948, Title: Foam Airless Spoked Tire -- FASTire (G.C.) The authors
acknowledge support from the Italian Ministry of Education, University and
Research (MIUR) under the program \textquotedblleft Departments of
Excellence\textquotedblright\ (L.232/2016)
\end{acknowledgments}

\bibliographystyle{plain}
\bibliography{paperviscoadh}

\providecommand{\noopsort}[1]{}\providecommand{\singleletter}[1]{#1}%
\begin{thebibliography}{10}

\bibitem{Baney1999}
J.~M. Baney and C.-Y. Hui.
\newblock Viscoelastic crack healing and adhesion.
\newblock {\em Journal of Applied Physics}, 86(8):4232--4241, October 1999.

\bibitem{Barquins1981}
M.~Barquins and D.~Maugis.
\newblock Tackiness of elastomers.
\newblock {\em The Journal of Adhesion}, 13(1):53--65, December 1981.

\bibitem{barquins1978}
M~Barquins, D~Maugis, J~Blouet, and R~Courtel.
\newblock Contact area of a ball rolling on an adhesive viscoelastic material.
\newblock {\em Wear}, 51(2):375--384, 1978.

\bibitem{Bhushan1995}
Bharat Bhushan, Jacob~N. Israelachvili, and Uzi Landman.
\newblock Nanotribology: friction, wear and lubrication at the atomic scale.
\newblock {\em Nature}, 374(6523):607--616, April 1995.

\bibitem{Carbone2004}
G~Carbone.
\newblock Adhesion and friction of an elastic half-space in contact with a
  slightly wavy rigid surface.
\newblock {\em Journal of the Mechanics and Physics of Solids},
  52(6):1267--1287, June 2004.

\bibitem{Carbone2005epje}
G.~Carbone and B.~N.~J. Persson.
\newblock Crack motion in viscoelastic solids: The role of the flash
  temperature.
\newblock {\em The European Physical Journal E}, 17(3):261--281, July 2005.

\bibitem{Carbone2005}
G.~Carbone and B.~N.~J. Persson.
\newblock Hot cracks in rubber: Origin of the giant toughness of rubberlike
  materials.
\newblock {\em Physical Review Letters}, 95(11), September 2005.

\bibitem{Carbone2013}
Giuseppe Carbone and Carmine Putignano.
\newblock A novel methodology to predict sliding and rolling friction of
  viscoelastic materials: Theory and experiments.
\newblock {\em Journal of the Mechanics and Physics of Solids},
  61(8):1822--1834, August 2013.

\bibitem{Charmet1996}
J.-C. Charmet and M.~Barquins.
\newblock Adhesive contact and rolling of a rigid cylinder under the pull of
  gravity on the underside of a smooth-surfaced sheet of rubber.
\newblock {\em International Journal of Adhesion and Adhesives},
  16(4):249--254, January 1996.

\bibitem{Chaudhury1996}
Manoj~K. Chaudhury.
\newblock Interfacial interaction between low-energy surfaces.
\newblock {\em Materials Science and Engineering: R: Reports}, 16(3):97--159,
  March 1996.

\bibitem{Chernyak1986}
Yu.B. Chernyak and A.I. Leonov.
\newblock On the theory of the adhesive friction of elastomers.
\newblock {\em Wear}, 108(2):105--138, March 1986.

\bibitem{LeGal2005}
Andr{\'{e}}~Le Gal, Xin Yang, and Manfred Kl\"{u}ppel.
\newblock Evaluation of sliding friction and contact mechanics of elastomers
  based on dynamic-mechanical analysis.
\newblock {\em The Journal of Chemical Physics}, 123(1):014704, July 2005.

\bibitem{Greenwood2004}
J~A Greenwood.
\newblock The theory of viscoelastic crack propagation and healing.
\newblock {\em Journal of Physics D: Applied Physics}, 37(18):2557--2569,
  September 2004.

\bibitem{Grosch1963}
K~A Grosch.
\newblock 274(1356):21--39, June 1963.

\bibitem{Hao2007}
S.~Hao and L.~M. Keer.
\newblock Rolling contact between rigid cylinder and semi-infinite elastic body
  with sliding and adhesion.
\newblock {\em Journal of Tribology}, 129(3):481--494, January 2007.

\bibitem{Hoyer2022}
Brodie~K. Hoyer, Rong Long, and Mark~E. Rentschler.
\newblock A tribometric device for the rolling contact of soft elastomers.
\newblock {\em Tribology Letters}, 70(2), March 2022.

\bibitem{Kim2007}
Seong~H. Kim, David~B. Asay, and Michael~T. Dugger.
\newblock Nanotribology and {MEMS}.
\newblock {\em Nano Today}, 2(5):22--29, October 2007.

\bibitem{Maeda2002}
Nobuo Maeda, Nianhuan Chen, Matthew Tirrell, and Jacob~N. Israelachvili.
\newblock Adhesion and friction mechanisms of polymer-on-polymer surfaces.
\newblock {\em Science}, 297(5580):379--382, July 2002.

\bibitem{Menga2016}
N.~Menga, L.~Afferrante, and G.~Carbone.
\newblock Effect of thickness and boundary conditions on the behavior of
  viscoelastic layers in sliding contact with wavy profiles.
\newblock {\em Journal of the Mechanics and Physics of Solids}, 95:517--529,
  October 2016.

\bibitem{Menga2018}
N.~Menga, L.~Afferrante, G.P. Demelio, and G.~Carbone.
\newblock Rough contact of sliding viscoelastic layers: numerical calculations
  and theoretical predictions.
\newblock {\em Tribology International}, 122:67--75, June 2018.

\bibitem{Menga2021}
N.~Menga, G.~Carbone, and D.~Dini.
\newblock Exploring the effect of geometric coupling on friction and energy
  dissipation in rough contacts of elastic and viscoelastic coatings.
\newblock {\em Journal of the Mechanics and Physics of Solids}, 148:104273,
  March 2021.

\bibitem{Menga2014}
N.~Menga, C.~Putignano, G.~Carbone, and G.~P. Demelio.
\newblock The sliding contact of a rigid wavy surface with a viscoelastic
  half-space.
\newblock {\em Proceedings of the Royal Society A: Mathematical, Physical and
  Engineering Sciences}, 470(2169):20140392, September 2014.

\bibitem{Persson2001}
B.~N.~J. Persson.
\newblock Theory of rubber friction and contact mechanics.
\newblock {\em The Journal of Chemical Physics}, 115(8):3840--3861, August
  2001.

\bibitem{Persson2017}
B.~N.~J. Persson.
\newblock Crack propagation in finite-sized viscoelastic solids with
  application to adhesion.
\newblock {\em {EPL} (Europhysics Letters)}, 119(1):18002, July 2017.

\bibitem{Persson2021}
B.~N.~J. Persson.
\newblock On opening crack propagation in viscoelastic solids.
\newblock {\em Tribology Letters}, 69(3), August 2021.

\bibitem{Persson2004}
B~N~J Persson, O~Albohr, U~Tartaglino, A~I Volokitin, and E~Tosatti.
\newblock On the nature of surface roughness with application to contact
  mechanics, sealing, rubber friction and adhesion.
\newblock {\em Journal of Physics: Condensed Matter}, 17(1):R1--R62, December
  2004.

\bibitem{Persson2005}
B.~N.~J. Persson and E.~A. Brener.
\newblock Crack propagation in viscoelastic solids.
\newblock {\em Physical Review E}, 71(3), March 2005.

\bibitem{roberts1979}
AD~Roberts.
\newblock Looking at rubber adhesion.
\newblock {\em Rubber Chemistry and Technology}, 52(1):23--42, 1979.

\bibitem{Scaraggi2015}
M~Scaraggi and B~N~J Persson.
\newblock Friction and universal contact area law for randomly rough
  viscoelastic contacts.
\newblock {\em Journal of Physics: Condensed Matter}, 27(10):105102, February
  2015.

\bibitem{Sharp2016}
R.~S. Sharp, P.~Gruber, and E.~Fina.
\newblock Circuit racing, track texture, temperature and rubber friction.
\newblock {\em Vehicle System Dynamics}, 54(4):510--525, January 2016.

\bibitem{She1998}
Hongquan She, David Malotky, and Manoj~K. Chaudhury.
\newblock Estimation of adhesion hysteresis at polymer/oxide interfaces using
  rolling contact mechanics.
\newblock {\em Langmuir}, 14(11):3090--3100, May 1998.

\bibitem{Shull2002}
Kenneth~R. Shull.
\newblock Contact mechanics and the adhesion of soft solids.
\newblock {\em Materials Science and Engineering: R: Reports}, 36(1):1--45,
  January 2002.

\bibitem{tiwari}
A.~Tiwari, L.~Dorogin, M.~Tahir, K.~W. Stöckelhuber, G.~Heinrich,
  N.~Espallargas, and B.~N.~J. Persson.
\newblock Rubber contact mechanics: adhesion, friction and leakage of seals.
\newblock {\em Soft Matter}, 13(48):9103--9121, 2017.

\bibitem{Vorvolakos2003}
Katherine Vorvolakos and Manoj~K. Chaudhury.
\newblock The effects of molecular weight and temperature on the kinetic
  friction of silicone rubbers.
\newblock {\em Langmuir}, 19(17):6778--6787, July 2003.

\bibitem{Yoshizawa1993_1}
Hisae Yoshizawa, You~Lung Chen, and Jacob Israelachvili.
\newblock Fundamental mechanisms of interfacial friction. 1. relation between
  adhesion and friction.
\newblock {\em The Journal of Physical Chemistry}, 97(16):4128--4140, April
  1993.

\bibitem{Yoshizawa1993_2}
Hisae Yoshizawa and Jacob Israelachvili.
\newblock Fundamental mechanisms of interfacial friction. 2. stick-slip
  friction of spherical and chain molecules.
\newblock {\em The Journal of Physical Chemistry}, 97(43):11300--11313, October
  1993.

\bibitem{Zhang2015}
Yuyan Zhang, Xiaoli Wang, Hanqing Li, and Weixu Yang.
\newblock A numerical study of the rolling friction between a microsphere and a
  substrate considering the adhesive effect.
\newblock {\em Journal of Physics D: Applied Physics}, 49(2):025501, November
  2015.

\bibitem{Zhou2013}
Ming Zhou, Noshir Pesika, Hongbo Zeng, Yu~Tian, and Jacob Israelachvili.
\newblock Recent advances in gecko adhesion and friction mechanisms and
  development of gecko-inspired dry adhesive surfaces.
\newblock {\em Friction}, 1(2):114--129, June 2013.

\end{thebibliography}

\end{document}